\documentclass[aps,prl,twocolumn,groupedaddress]{revtex4}

\usepackage{graphicx}
\usepackage{bm}
\usepackage{amsmath}
\usepackage{subfigure}
\begin{document}

\title{Chirality and Biaxiality in Cholesteric Liquid Crystals}

\author{Subas Dhakal}
\author{Jonathan V. Selinger}
\email{jselinge@kent.edu}
\affiliation{Liquid Crystal Institute, Kent State University, Kent, OH 44242}

\date{June 23, 2010}

\begin{abstract}
We investigate the statistical mechanics of chirality and biaxiality in liquid
crystals through a variety of theoretical approaches, including Monte Carlo
simulations, lattice mean-field theory, and Landau theory.  All of these
calculations show that there is an important interaction between cholesteric
twist and biaxial order:  The twist acts as a field on the biaxial order, and
conversely, the biaxial order increases the twist, i.e. reduces the pitch.
We model the behavior of chiral biaxial liquid crystals as a function of
temperature, and discuss how the predictions can be tested in experiments.
\end{abstract}


\maketitle

In liquid crystals, there is a close connection between \emph{chirality},
asymmetry under reflection, and \emph{biaxiality}, orientational order in the
plane perpendicular to the director.  In the 1970s, Priest and
Lubensky~\cite{priest74} recognized that a cholesteric liquid crystal must
have some slight biaxial order because of the difference between the
directions along and perpendicular to the helical axis.  Brand and
Pleiner~\cite{brand85} showed theoretically that chirality can smear out the
transition between uniaxial and biaxial phases, and Kroin
et al.~\cite{kroin89} confirmed this smearing experimentally in lyotropic
liquid crystals.  Later, Harris, Kamien, and Lubensky~\cite{harris97}
developed a microscopic model of molecules interacting through classical
central-force interatomic potentials, and found that cholesteric twist can
only form if there are at least short-range biaxial correlations between
molecules.  By comparison, in a system with quantum dispersive interactions,
cholesteric twist can form even without such correlations~\cite{issaenko99}.

In recent years, there has been a resurgence of interest in biaxial liquid
crystals---driven in part by experimental reports of the discovery of a
biaxial nematic phase in thermotropic liquid
crystals~\cite{acharya04,madsen04}, and in part by prospects for using biaxial
liquid crystals for fast-switching display devices~\cite{kumar09}.  For that
reason, it is now important to re-examine the interplay between chirality and
biaxiality in liquid crystals.  The key issue is:  How is the cholesteric
pitch affected by biaxial order---either by long-range biaxial order or by
short-range biaxial correlations?

In this paper, we investigate this issue through three theoretical approaches:
(1)~Monte Carlo simulations of a lattice model for chiral molecules
interacting via anisotropic van der Waals forces.  (2)~Mean-field theory for
the same lattice model.  (3)~Landau theory based on symmetry-allowed couplings
between twist and biaxial order.  Through all three approaches, we calculate
the cholesteric twist as a function of molecular chirality, molecular
biaxiality, and temperature.  These calculations show that chirality acts as
an effective field on the biaxial order, which changes the second-order
uniaxial-biaxial transition into a rapid but nonsingular evolution.
Conversely, biaxial order enhances the cholesteric twist, i.e. reduces the
pitch, so that the pitch greatly decreases in the low-temperature, highly
biaxial state.  The calculations also allow us to reconsider the relationship
between twist and short-range biaxial correlations.  Based on this theoretical
work, we discuss opportunities for experimental studies of chiral biaxial
liquid crystals.

\begin{figure}[b]
(a)\subfigure{\includegraphics[width=0.75in,clip=true]{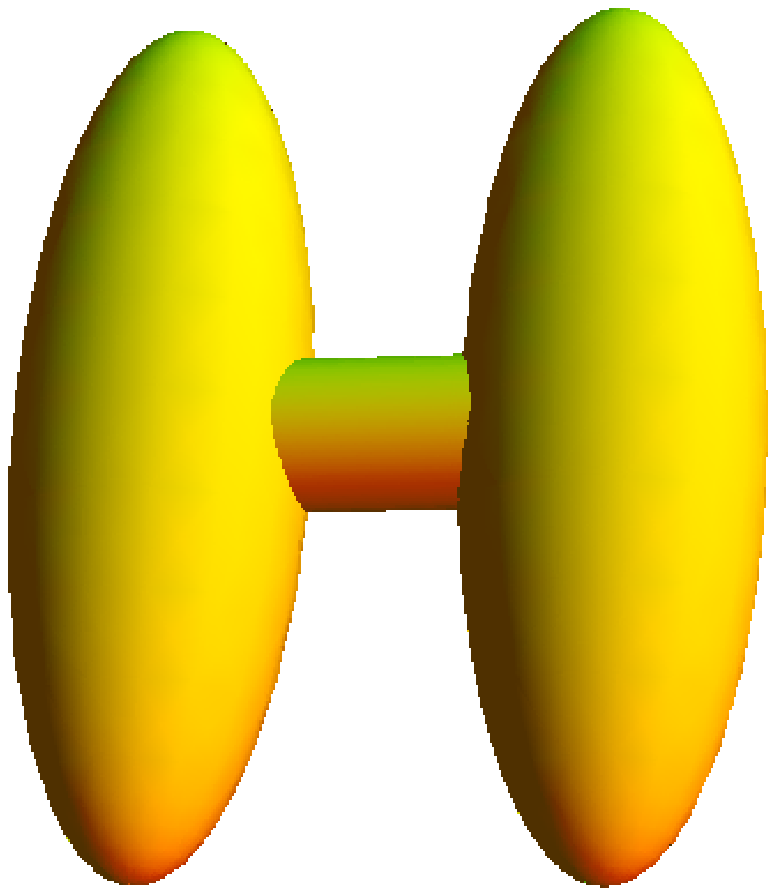}}
(b)\subfigure{\includegraphics[width=0.75in,clip=true]{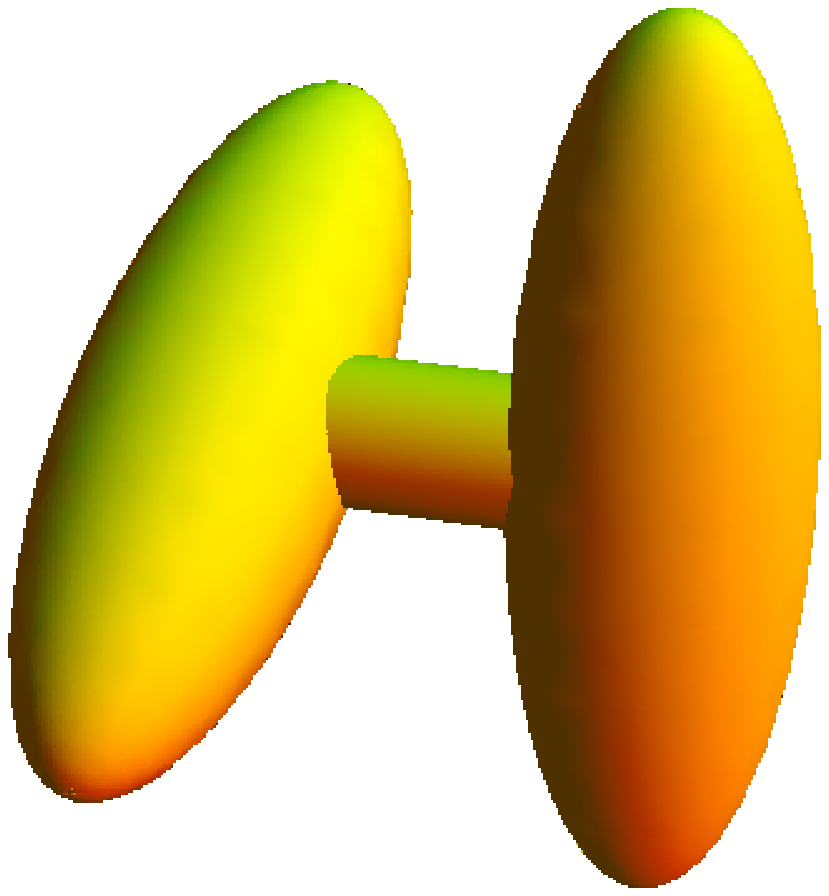}}
(c)\subfigure{\includegraphics[width=3.0in,clip=true]{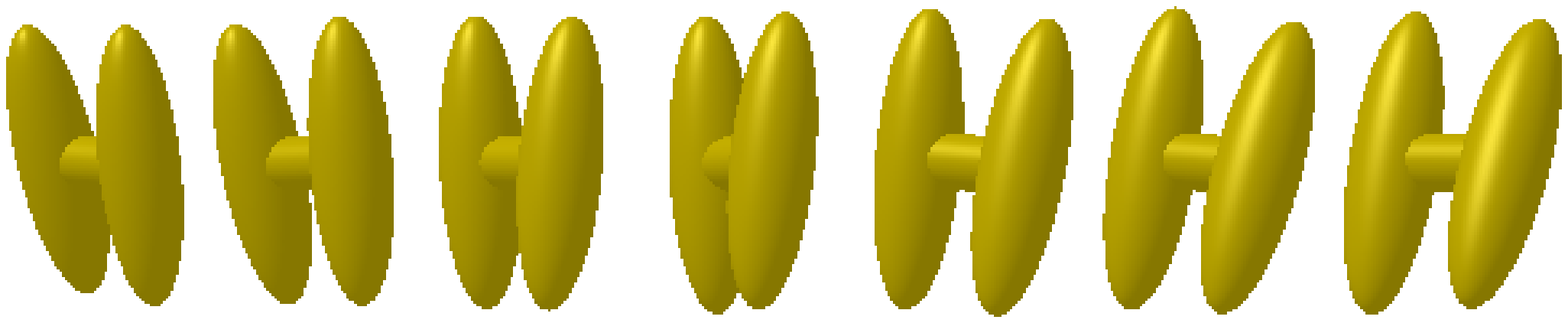}}
\caption{(a)~Achiral and (b)~chiral biaxial molecular structures studied in
this work.  (c)~Cholesteric phase of chiral molecules, showing the macroscopic
twist.}
\end{figure}

For the simulations, we need a model molecular structure that can exhibit
biaxial order with or without chirality.  Inspired by van der Meer
et al.~\cite{vandermeer76}, we consider a structure with two ellipsoids
arranged rigidly in the shape of the letter~H.  Each ellipsoid represents an
extended, anisotropic charge distribution within the molecule.  If the two
ellipsoids are parallel, as in Fig.~1(a), this is an achiral biaxial
structure, with a biaxiality characterized by the separation $h$.
By contrast, if the ellipsoids are twisted about the central connector, as in
Fig.~1(b), this is a chiral biaxial structure, with a chirality characterized
by the twist angle $\chi$ of each ellipsoid from the parallel configuration.
The interaction between any two ellipsoids on
neighboring molecules is the van der Waals dipole-induced-dipole interaction.
Hence, the total interaction between two molecules $i$ and $j$ is the sum of
four pairwise interactions among the constituent ellipsoids,
\begin{equation}
U_{ij}=-A\sum_{\alpha,\beta=1,2}
\frac{(\hat{\bm{e}}_{i\alpha}\cdot\hat{\bm{e}}_{j\beta})^2}%
{r_{i\alpha,j\beta}^6},
\label{interaction}
\end{equation}
where $\hat{\bm{e}}_{i\alpha}$ is the orientation of ellipsoid $\alpha$ on
molecule $i$, and $r_{i\alpha,j\beta}=|\bm{r}_{j\beta}-\bm{r}_{i\alpha}|$ is
the center-to-center distance between two interacting ellipsoids $i\alpha$ and
$j\beta$.

\begin{figure*}
(a)\subfigure{\includegraphics[width=3in]{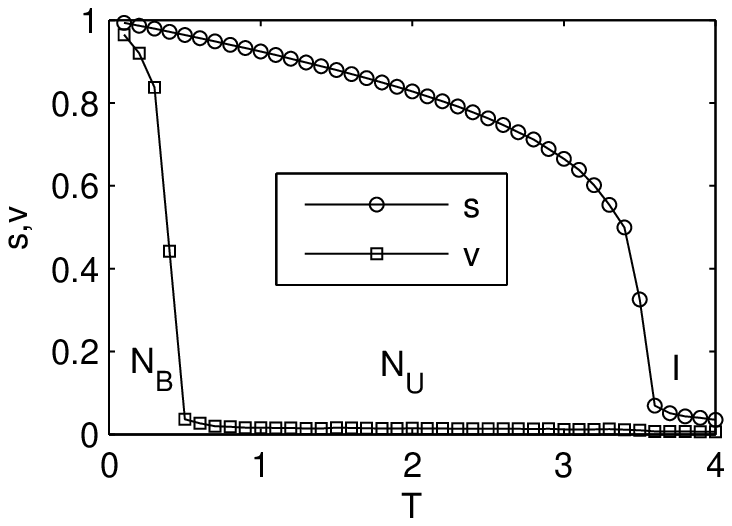}}
(b)\subfigure{\includegraphics[width=3in]{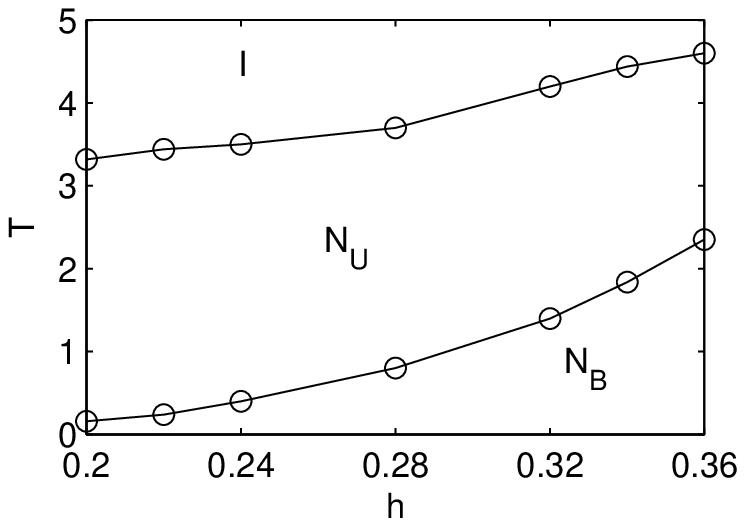}}
(c)\subfigure{\includegraphics[width=3in]{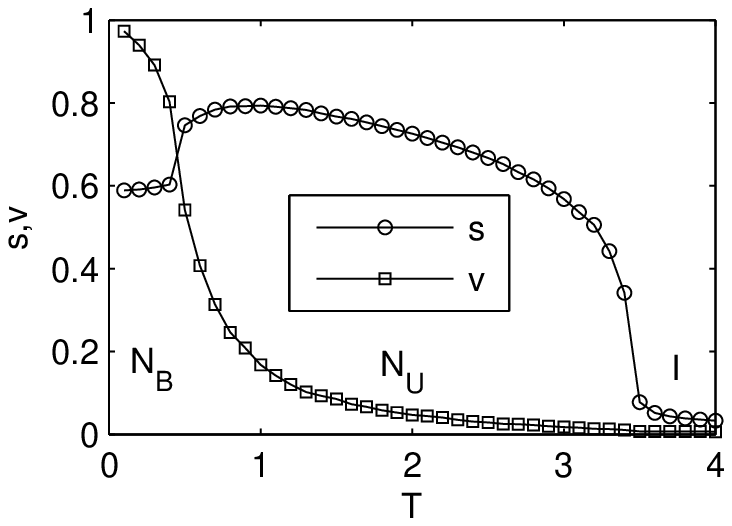}}
(d)\subfigure{\includegraphics[width=3in]{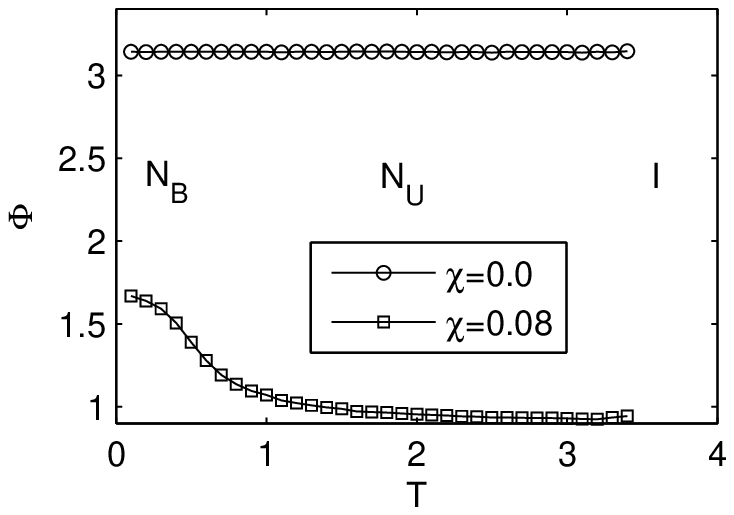}}
\caption{Monte Carlo simulation results: (a)~Uniaxial and biaxial order
parameters as functions of temperature $T$, for achiral biaxial molecules with
ellipsoid separation $h=0.24$.  (b)~Complete phase diagram for achiral biaxial
molecules, in terms of $h$ and~$T$.  (c)~Uniaxial and biaxial order parameters
as functions of $T$, for chiral biaxial molecules with $h=0.24$ and molecular
twist angle $\chi=0.08$.  (d)~Boundary twist angle $\Phi$ as a function of
$T$.  For achiral ($\chi=0$) molecules, $\Phi$ is locked at $\pi$, indicating
that the system is not twisted.  For chiral ($\chi=0.08$) molecules, $\Phi$ is
not a multiple of $\pi$, indicating that the system is twisted, and the
cholesteric twist increases as $T$ decreases.}
\end{figure*}

Initially, we perform Monte Carlo simulations of \emph{achiral} biaxial
molecules.  We simulate a simple cubic lattice of size $16\times16\times16$,
with a molecule centered on each lattice site.  In each Monte Carlo step, a
molecule is randomly selected and its orientation is changed, following the
standard Metropolis algorithm.  The uniaxial order parameter $S$ and biaxial
order parameter $V$ are calculated as described by Bates and
Luckhurst~\cite{bates05}.  For small ellipsoid separation $h$, we find a
first-order transition from isotropic~($I$) to uniaxial nematic~($N_u$),
followed by a second-order transition to biaxial nematic~($N_b$) at lower
temperature.  The temperature range of the biaxial nematic phase increases
with $h$, as expected for board-shaped molecules.  Figure~2(a) shows a
sample plot of the order parameters for $h=0.24$, and Fig.~2(b) shows the full
simulated phase diagram.

We now use the same approach to simulate \emph{chiral} biaxial molecules.  In
this system, we expect molecular chirality to induce a cholesteric twist.
This twist is generally not consistent with periodic boundary conditions.
Hence, we use self-adjusting twisted boundary conditions in the $z$-direction,
following the method of Memmer~\cite{memmer01}.  In this method, the boundary
twist angle $\Phi$ from the top to bottom of the cell is a free simulation
variable, determined by the Monte Carlo process. In an untwisted system,
$\Phi$ must be a multiple of $\pi$.  Hence, the deviation of $\Phi$ from a
multiple of $\pi$ is a measure of the twist across the system, i.e. the
inverse pitch.  With this method, the simulation forms a cholesteric phase
over a wide temperature range.  A sample configuration showing the molecular
orientations along the $z$-axis is shown in Fig.~1(c).

Using these simulations, we determine the uniaxial and biaxial order
parameters for systems of chiral molecules.  Figure~2(c) shows $S$ and $V$ as
functions of temperature $T$ for ellipsoid separation $h=0.24$ and molecular
twist angle $\chi=0.08$.  At $T=3.4$ there is a first-order transition from
isotropic to cholesteric, as seen from the jump in $S$. In the cholesteric
phase there is a slight nonzero value of $V$, as expected from
Ref.~\cite{priest74}.  As $T$ decreases further, $V$ gradually increases
toward its maximum value of $1$.  There is no phase transition between
uniaxial and biaxial, but only a nonsingular increase in $V$.  Apparently the
chirality acts as an effective field on the biaxial order, which smears out
the $N_u$-$N_b$ transition.

We also determine the boundary twist angle $\Phi$ as a function of $T$, as
shown in Fig.~2(d). For achiral molecules, the boundary twist angle is locked
at $\Phi=\pi$, indicating that the system is in a uniform nematic phase,
either uniaxial or biaxial.  By contrast, for chiral molecules with
$\chi=0.08$, the results for $\Phi$ show a twisted cholesteric phase.  The
cholesteric twist is substantial just below the isotropic-cholesteric
transition, although the biaxial order parameter is very small there.  The
twist increases further as $T$ decreases, and becomes largest in the
temperature range that \emph{would be} the biaxial phase for an achiral
system.  Thus, we see that the cholesteric twist and the biaxial order
increase together, reinforcing each other, as $T$ decreases.

In addition to the long-range biaxial order, we measure the short-range
biaxial correlations as a function of distance between nearby lattice sites.
Through most of the cholesteric temperature range, these biaxial correlations
are very small; they do not become noticeable until slightly above the achiral
$N_u$-$N_b$ transition temperature.

To compare with the simulations, we construct a Maier-Saupe-type mean-field
theory for the same lattice model, following a method similar to our
calculation for the flexoelectric effect~\cite{dhakal10}.  Here, we assume the
system has perfect order of the long axes of the molecules, but variable
biaxial order and variable cholesteric twist.  Suppose that site $i$ has its
long axis along the $x$-direction, as do the four neighbors in the $xy$-plane,
while the two neighbors in the $z$-direction have long axes twisted about the
$z$-axis. The long axis at site~$i$ is $\hat{\bm{n}}_i=(1,0,0)$, while the
long axes of the neighbors are $\hat{\bm{n}}_{\pm x}=(1,0,0)$,
$\hat{\bm{n}}_{\pm y}=(1,0,0)$, and
$\hat{\bm{n}}_{\pm z}=R_z(\pm\Delta\theta)(1,0,0)$, where $\Delta\theta$ is
the cholesteric twist from one layer to the next and $R_z(\Delta\theta)$ is
the rotation operator about the $z$-axis.  At each site, the molecular short
axis must be in the plane perpendicular to the long axis.  Hence, the short
axis at site $i$ is $\hat{\bm{b}}_i=(0,\sin\phi_i,\cos\phi_i)$, while the
short axes of the neighbors are
$\hat{\bm{b}}_{\pm x}=(0,\sin\phi_{\pm x},\cos\phi_{\pm x})$,
$\hat{\bm{b}}_{\pm y}=(0,\sin\phi_{\pm y},\cos\phi_{\pm y})$, and
$\hat{\bm{b}}_{\pm z}=
R_z(\pm\Delta\theta)(0,\sin\phi_{\pm z},\cos\phi_{\pm z})$, where the local
angle $\phi$ represents the azimuthal angle of the short axis.

We now construct a distribution function for the local azimuthal angle $\phi$,
which can be written as
\begin{equation}
\rho(\phi)=\frac{\exp(C\cos 2\phi)}{\int_0^{2\pi}d\phi\exp(C\cos 2\phi)},
\end{equation}
where $C$ is a variational parameter representing the effective biaxial
potential.  It is related to the biaxial order parameter by
$V=\int_0^{2\pi}\cos(2\phi)\rho(\phi)d\phi$.
With this distribution function, the mean-field free energy per site can be
written as the sum of energetic and entropic terms,
\begin{equation}
F=\langle H\rangle+k_B T\langle\log\rho\rangle,
\label{Fmeanfield}
\end{equation}
where $\langle H\rangle$ is the average interaction energy of
Eq.~(\ref{interaction}) between site $i$ and its six neighbors.  This free
energy depends on two variational parameters (effective biaxial potential $C$
and cholesteric twist $\Delta\theta$), two molecular parameters (ellipsoid
separation $h$ and molecular chirality $\chi$), and temperature $T$ (in units
of interaction strength $A$).  We numerically minimize the free energy over
the variational parameters for each set of molecular parameters and
temperature, to find the biaxial order and twist.

\begin{figure}
\includegraphics[width=3in]{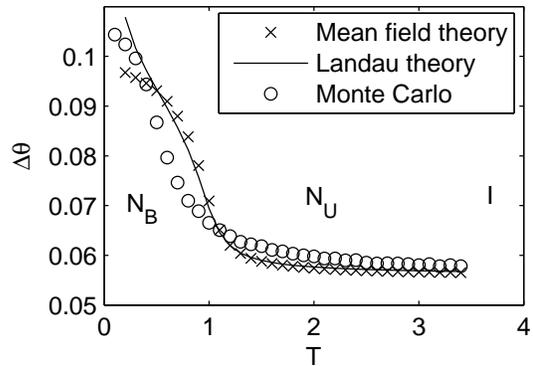}
\caption{Theoretical results for cholesteric twist as a
function of temperature (in units of interaction strength $A$):  Monte Carlo
simulations, mean-field theory, and Landau theory.}
\end{figure}

The numerical mean-field results are consistent with the Monte Carlo
simulations.  For \emph{achiral} biaxial molecules, the system has a uniaxial
phase with $V=0$ at high temperature.  As a critical temperature, it undergoes
a second-order transition to a biaxial phase with $V\neq0$, and the biaxial
order parameter increases as a power law as $T$ decreases.  This achiral
system is untwisted, with $\Delta\theta=0$ for all $T$.  By contrast, for
\emph{chiral} biaxial molecules, the system has a high-temperature cholesteric
phase with a small but nonzero value of $V$.  As $T$ decreases, $V$ increases
gradually, without any phase transition between uniaxial and biaxial.  The
cholesteric twist $\Delta\theta$ is substantial at high temperature, even when
$V$ is small, and it increases further as $T$ decreases, as shown
by crosses in Fig.~3.  Thus, as in the simulations, we see that the chirality
acts as a field that induces biaxial order and smears out the $N_u$-$N_b$
transition, and conversely, the biaxial order increases the cholesteric twist.

For further insight into the relationship between cholesteric twist and
biaxial order, we construct a Landau theory for a chiral biaxial liquid
crystal.  As in the mean-field theory above, we suppose the system has perfect
uniaxial order along the local axis $\hat{\bm{n}}(\bm{r})$, but variable
biaxial order.  The biaxial order can be described by the tensor
$B_{ij}=V(b_i b_j - c_i c_j)$, where $V$ is the magnitude of the order, and
$\hat{\bm{b}}(\bm{r})$ and $\hat{\bm{c}}(\bm{r})$ are the two principal axes
orthogonal to $\hat{\bm{n}}(\bm{r})$.  The free energy can then be expanded in
$B_{ij}$ and in gradients of $\hat{\bm{n}}(\bm{r})$, to obtain
\begin{eqnarray}
F&=&\textstyle{\frac{1}{2}}K(\partial_i n_j)(\partial_i n_j)
    -K q_0\epsilon_{ijk}n_i \partial_j n_k\nonumber\\
&&+\textstyle{\frac{1}{4}}r(T-T_\mathrm{UB})\mbox{Tr}(B^2)
  +\textstyle{\frac{1}{8}}s\mbox{Tr}(B^4)
  +\textstyle{\frac{1}{12}}t\mbox{Tr}(B^6)\nonumber\\
&&-u\epsilon_{ijk}B_{jl}n_i \partial_l n_k
  -w\epsilon_{ijk}B_{jl}n_i \partial_k n_l
\end{eqnarray}
In this expression, the first line is the Frank free energy for director
gradients in a chiral liquid crystal, the second line is a power series
expansion in $B_{ij}$, and the third line is a pair of chiral couplings
between $B_{ij}$ and director gradients.  If we now assume a cholesteric
modulation of the form $\hat{\bm{n}}=(\cos q z,\sin q z,0)$,
$\hat{\bm{b}}=(0,0,1)$, and $\hat{\bm{c}}=(-\sin q z,\cos q z,0)$, with an
arbitrary twist wave vector $q$, the free energy simplifies to
\begin{eqnarray}
F&=&\textstyle{\frac{1}{2}}K q^2
-K q_0 q
+\textstyle{\frac{1}{2}}r(T-T_\mathrm{UB})V^2
+\textstyle{\frac{1}{4}}s V^4
+\textstyle{\frac{1}{6}}t V^6\nonumber\\
&&-(u+w)V q.
\end{eqnarray}
In the limit of high temperature, where biaxial order is small and the $s$ and
$t$ terms are negligible, we minimize this free energy over $V$ and $q$ to
obtain
\begin{subequations}
\label{landausolutions}
\begin{eqnarray}
V &\approx& \frac{(u+w)q_0}{r(T-T_\mathrm{UB})},\\
q &\approx& q_0 + \frac{(u+w)^2 q_0}{K r(T-T_\mathrm{UB})}.
\end{eqnarray}
\end{subequations}
Equations~(\ref{landausolutions}) demonstrate that the twist~$q$ acts as a
field on the biaxial order~$V$, and conversely, the biaxial order increases
the twist, and hence reduces the pitch.

Instead of treating the Landau coefficients as purely phenomenological
parameters, we can derive them from the lattice model presented in this paper,
by expanding the free energy of Eq.~(\ref{Fmeanfield}) in powers of biaxial
order and twist.  Results of this calculation are shown by the solid line in
Fig.~3.  The predictions of Landau theory are consistent with simulation and
mean-field results, except at low temperature where biaxial order is large and
series expansion is unreliable.

It is interesting to compare our results with Ref.~\cite{harris97}, which
argued that short-range biaxial correlations are a key factor in determining
cholesteric twist in systems with classical central-force interactions.  We
also find an important connection between biaxiality and cholesteric twist,
but it differs from their argument in two ways:

(a)~Our model shows some twist even in the limit of no biaxiality.  This
result does not contradict Ref.~\cite{harris97}, because our system does not
have central-force interactions; it is consistent with Ref.~\cite{issaenko99}
for quantum dispersive interactions.  However, it draws attention to the fact
that most liquid crystals have quantum dispersive interactions, while
central-force interactions are unusual.  (To be sure, quantum dispersive
interactions are derived from fluctuating microscopic central-force
interactions among electrons, and these electrons might have some biaxial
correlations.  However, such correlations would be difficult to observe in
either experiments or simulations; normal observations average over the
fluctuations.)

(b)~Our results show there is not a cause-and-effect relationship between
biaxiality and cholesteric twist; rather, there is a mutually reinforcing
interaction between them.  Twist acts as a field on biaxial order, and
conversely, biaxial order helps to increase twist.  As temperature decreases,
biaxiality and twist increase together.

The theory presented here has implications for experiments on biaxial liquid
crystals, either thermotropic or lyotropic.  It should be possible to choose
an achiral host that has a uniaxial-biaxial transition, and add a chiral
dopant.  The theory predicts that the dopant will induce a small twist (large
pitch) in the uniaxial phase, but the twist will increase (pitch will
decrease) as the uniaxial-biaxial transition is approached.  For low dopant
concentration the twist will diverge as $(T-T_\mathrm{UB})^{-1}$, while for
larger concentration the divergence will be more rounded.  At the same time,
the chiral dopant will smear out the uniaxial-biaxial transition.  The
chirality-induced rounding of the uniaxial-biaxial transition has been
observed in lyotropics~\cite{kroin89}, but has not yet been investigated in
thermotropics.  Moreover, to our knowledge, no experiments have yet examined
the cholesteric twist around the uniaxial-biaxial transition in either
thermotropics or lyotropics.  This should be a promising area for experimental
research, to further characterize the close relationship between chirality and
biaxiality.

We thank D.~W.~Allender, R.~D. Kamien, and R.~L.~B. Selinger for helpful
discussions.  This work was supported by NSF Grant DMR-0605889.  Computational
resources were provided by the Ohio Supercomputer Center and the Wright Center
of Innovation for Advanced Data Management and Analysis.

\end{document}